\documentclass[a4paper]{jpconf}
\pdfoutput=1
\usepackage{graphicx}
\def \ket|#1>{\left|{#1}\right\rangle}
\def \bra<#1|{\left\langle{#1}\right|}
\def \braket<#1>{\langle{#1}\rangle}

\def \({\left(}
\def \){\right)}
\def \_#1{_{\rm #1}}
\def \^#1{^{\rm #1}}
\def \->{\rightarrow}
\def \bm #1{\mbox{\boldmath{${#1}$}}}

\begin{document}
\title{Transport properties in network models with perfectly conducting channels}

\author{K Kobayashi$^1$, K Hirose$^1$, H Obuse$^2$, T Ohtsuki$^1$ and K Slevin$^3$}

\address{$^1$
 Department of Physics, Sophia University, 102-8554 Tokyo, Japan
}\address{$^2$
 Department of Physics, Kyoto University,  606-8502 Kyoto, Japan
}\address{$^3$
 Department of Physics, Osaka University,  560-0043 Osaka, Japan
}

\ead{k-koji@sophia.ac.jp}

\begin{abstract}
 We study the transport properties of disordered electron systems that contain perfectly conducting channels.
 Two quantum network models that belong to different universality classes, unitary and symplectic, are simulated numerically.
 The perfectly conducting channel in the unitary class can be realized in zigzag graphene nano-ribbons and that in the symplectic class is known to appear in metallic carbon nanotubes.
 The existence of a perfectly conducting channel leads to novel conductance distribution functions and a shortening of the conductance decay length.
\end{abstract}

\section{Introduction}
 The Chalker-Coddington network model \cite{CC,KOK} (C-C model) is one of the most useful models for describing quantum transport in disordered two-dimensional (2D) electron systems in strong perpendicular magnetic fields.
 The C-C model can be extended to describe a system with one perfectly conducting channel (PCC) \cite{Hirose}.
 The PCC is a channel whose transmission eigenvalue is one, regardless of the phase of the system.
 The PCC in a strong magnetic field can be realized in zigzag graphene nano-ribbons \cite{Wakabayashi}.
 
 Although the original C-C model belongs to the unitary universality class, it is possible to construct network models that belong to the symplectic class \cite{Obuse}.
 The symplectic network model can also be extended to have a PCC of a Kramers doublet.
 The PCC in the symplectic class appears in metallic carbon nanotubes \cite{AndoSuzuura}.

 In this paper, we study 
 how a PCC affects the transport properties in the unitary and symplectic class.
 Though the models and origins of the PCC are completely different for these two cases, we show that the effects of the PCC are similar.

\section{Models}
 Transport in a network model is described by a scattering matrix $\bm S$,
	\begin{equation}
		\left(
		\begin{array}{@{\,}c@{\,}}
			\psi\^{out}\_{L}  \\
			\psi\^{out}\_{R} 
		\end{array}
		\right)
		=\bm{S}
		\left(
		\begin{array}{@{\,}cc@{\,}}
			\psi\^{in}\_{L}  \\
			\psi\^{in}\_{R} 
		\end{array}
		\right),
		\quad
		\bm{S}=
		\left(
		\begin{array}{@{\,}cc@{\,}}
			\bm{r} & \bm{t}' \\
			\bm{t} & \bm{r}'
		\end{array}
		\right),
	\end{equation}
where $\psi\^{in/out}\_{L/R}$ denotes the incoming/outgoing current amplitude on the left/right terminal of the network.
 The dimensionless conductance $G$ is given by
	\begin{equation}
		G = \Tr(\bm t^{\dagger}\bm t).
	\end{equation}
 We have calculated $\bm t$ using the method described in Ref. \cite{KOK}.

\subsection{Network model with PCC in the unitary class}
\begin{figure}[h] 
 \begin{minipage}[t]{75mm}
  \begin{center}
 \includegraphics[width=60mm, bb=0 0 575 407]{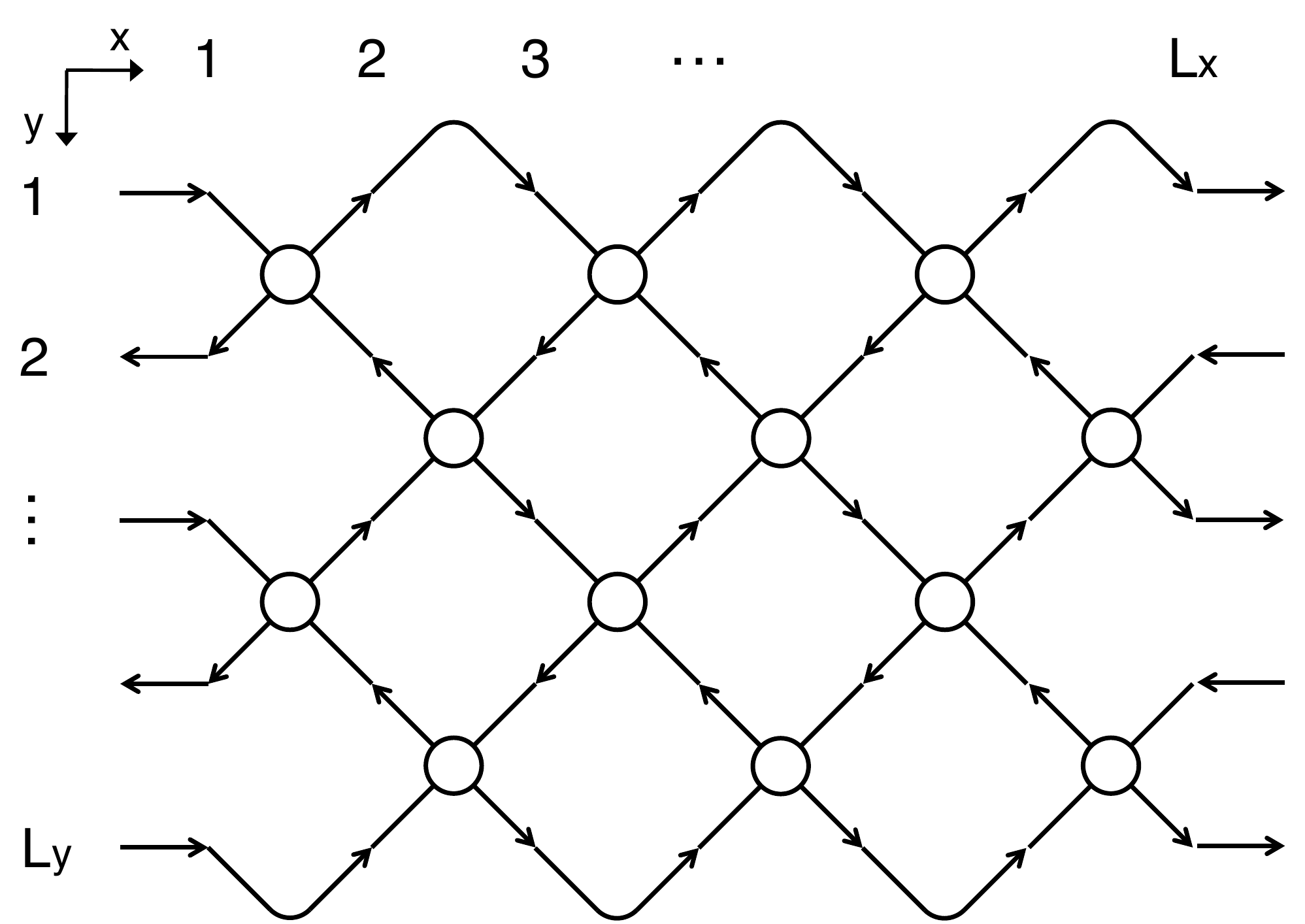}\hspace{8.4mm}%
  \caption{Schematic diagram of a unitary network model with a PCC.
           The electrons move along the solid lines in the directions indicated by the arrows and are scattered at the nodes.}
  \label{Unet}
  \end{center}
 \end{minipage}\hspace{8mm}
 \begin{minipage}[t]{75mm}
  \begin{center}
 \includegraphics[width=60mm, bb=0 0 583 412]{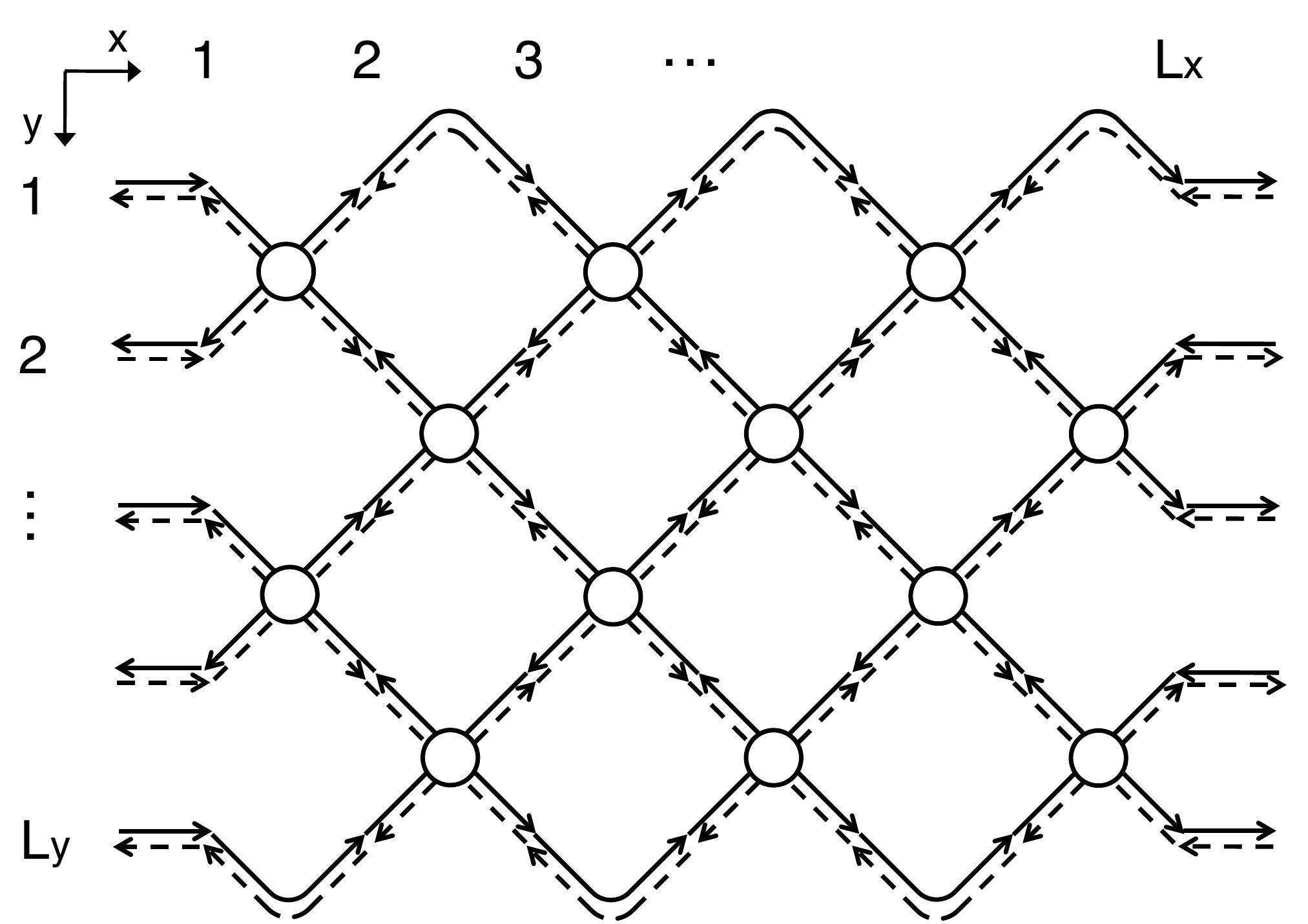}\hspace{8.4mm}%
  \caption{Schematic diagram of a symplectic network model with a PCC.
           The up (down) spin electrons move along the solid (dashed) lines in the directions indicated by the arrows and are scattered at the nodes.}
  \label{Snet}
  \end{center}
 \end{minipage}
\end{figure}

 The unitary network model with a PCC (Figure \ref{Unet}) is constructed by adding one extra channel to the original C-C model.
 The new model has $L_y=2N+1$ channels ($N+1$ right-directed and $N$ left-directed channels).
 In the unitary case, a PCC arises from the imbalance in the number of the incoming and outgoing channels \cite{Hirose,Barnes}, leading to a rectangular reflection matrix.

\subsection{Network model with a PCC of a Kramers doublet in the symplectic class}
 The symplectic network model with a PCC (Figure \ref{Snet}) is constructed by adding one extra channel of a Kramers doublet (i.e., two extra channels of up and down spins moving in opposite directions) to the symplectic network model proposed in Ref. \cite{Obuse}.
 The new model has $2L_y=4N+2$ channels ($2N+1$ right-directed and $2N+1$ left-directed channels).
 In this case, the reflection matrix is an antisymmetric square matrix with odd dimensions.
 This gives rise to the PCC \cite{AndoSuzuura}.

 In all cases, we have imposed the fixed boundary conditions (FBC) in the transverse direction.

\section{Results} 
\subsection{Critical conductance in 2D systems}
 Conductance distributions in disordered electron systems are universal at the Anderson transition points.
 In the 2D unitary class, the network has one transition point at the integer quantum Hall transition, 
 irrespective of the existence of the PCC (number of the PCC $N\_{PCC}=0,1$).
 In the symplectic class, the network without a PCC (with FBC) has two types of transitions \cite{ObuseNew}, the metal-insulator and metal-insulator with two Kramers pairs of edge states (i.e. topological insulator) transition.
 In this paper, we are not concerned with the latter.
 On the other hand, the symplectic network with a PCC has only the metal-insulator with one Kramers pair of edge states transition \cite{ObuseNew}.

\begin{figure}[h] 
 \begin{minipage}[t]{75mm}
  \begin{center}
   \includegraphics[width=64mm, bb=0 0 342 257]{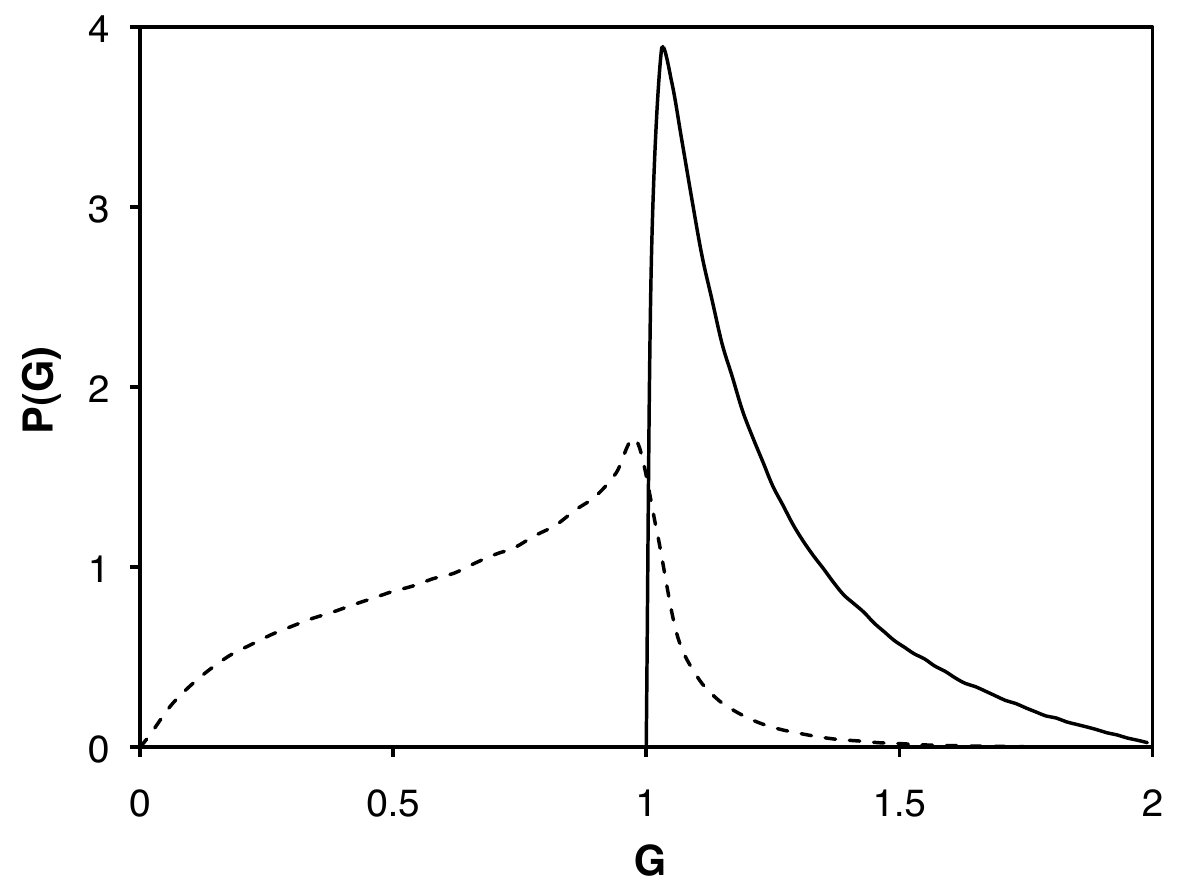}
  \caption{Distribution of the conductance at transition points in the unitary class
           with a PCC at $L_x=L_y=129$ (\full) 
           and without a PCC at $L_x=L_y=128$ (\dashed). 
           The size of the system is twice as large as in the case of Ref. \cite{Hirose}.
           }
  \label{P(G)_U}
  \end{center}
 \end{minipage}\hspace{8mm}
 \begin{minipage}[t]{75mm}
  \begin{center}
   \includegraphics[width=64mm, bb=0 0 342 257]{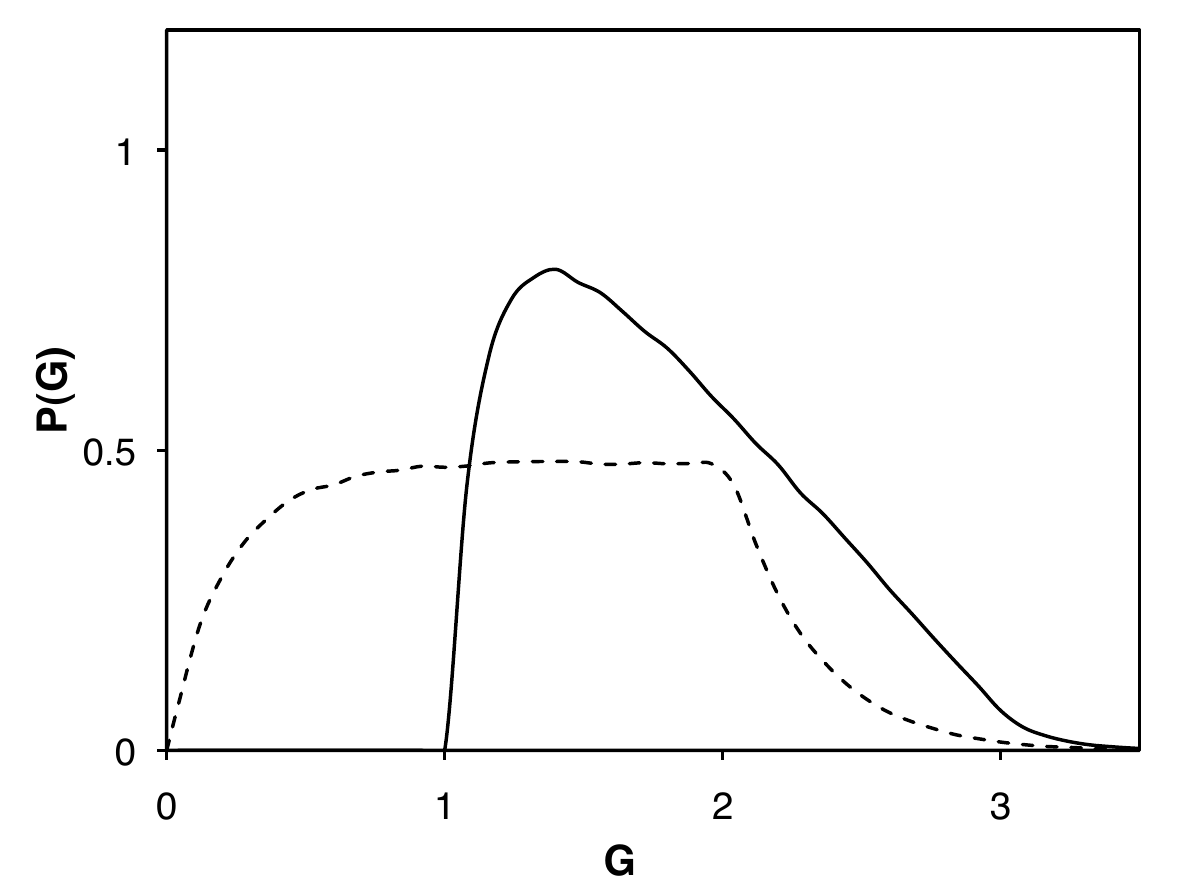}
  \caption{Distribution of the conductance at transition points in the symplectic class
           with a PCC at $L_x=L_y=129$ (\full) 
           and without a PCC at $L_x=L_y=128$ (\dashed). 
           }
  \label{P(G)_S}
  \end{center}
 \end{minipage}
\end{figure}

 The conductance distribution functions $P(G)$ calculated for the unitary and the symplectic class in the square geometry are shown in Figures \ref{P(G)_U} and \ref{P(G)_S}, respectively.
 The ensemble averaged conductance $\braket<G\^{2D}>$ and its variance ${\rm var}(G\^{2D})$ 
at criticality are listed in Table \ref{G_2D}.
 Since the PCC raises the minimum conductance to one, 
the conductance for the network models with a PCC is always larger than one.
 The distribution $P(G)$ for network models with a PCC differs completely from that for networks without a PCC.
 For systems with a PCC, the conductance distribution rises sharply from $G=1$.
 We also see that the variances for both models are reduced almost by one half when the PCC is added.
 Note that the distribution $P(G)$ for the models with the PCC is still universal.

\begin{table}[h] 
 \caption{Critical conductance in 2D.}
 \label{G_2D}
 \begin{center}
  \begin{tabular}{rccc}
   \br
    universality class & $N\_{PCC}$ & $\braket<G\^{2D}>$ & ${\rm var\,}(G\^{2D})$\\
   \mr
    unitary    \ & 1 &  1.23  &  0.043 \\
    unitary    \ & 0 &  0.69  &  0.087 \\[1mm]
    symplectic \ & 1 &  1.81  &  0.24 \\
    symplectic \ & 0 &  1.27  &  0.43 \\
   \br
  \end{tabular}
 \end{center}
\end{table}

\subsection{Decay of conductance in the quasi-one dimensional (Q1D) quantum wire}
 Though the PCC enhances the conductance, the transmission eigenvalue `1' of the PCC repels other transmission eigenvalues \cite{TakaneDMPK}.
 In other words, the PCC makes the length scale of the conductance decay in a quantum wire shorter.

\begin{figure}[h] 
 \begin{minipage}[t]{75mm}
   \includegraphics[width=72mm, bb=0 0 399 240]{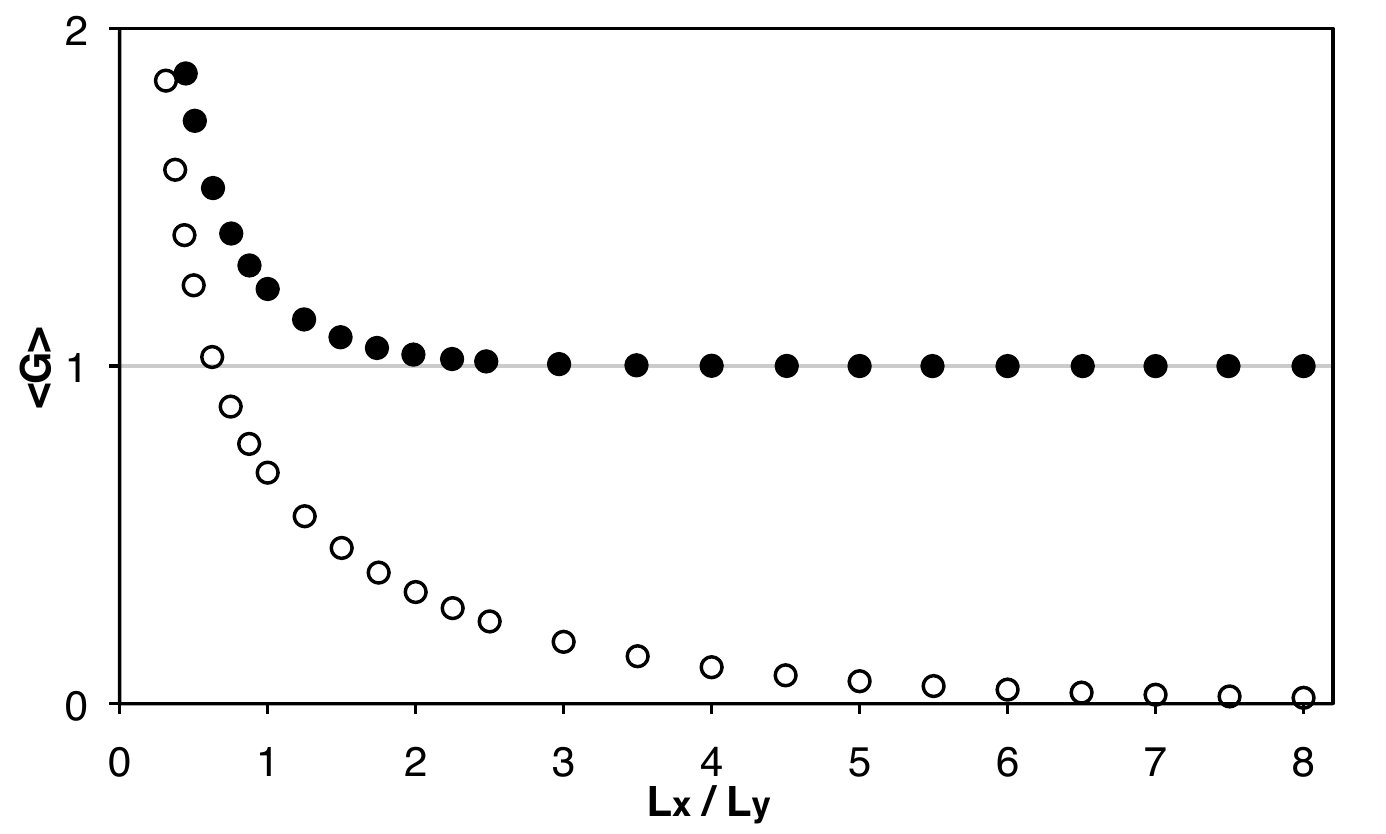}
  \caption{System length dependence of the average conductances at transition points in the unitary class
           with a PCC at $L_y=65$ ($\bullet$)
           and without a PCC at $L_y=64$ ($\circ$).
           }
  \label{x-dep_U}
 \end{minipage}\hspace{8mm}%
 \begin{minipage}[t]{75mm}
   \includegraphics[width=72mm, bb=0 0 399 238]{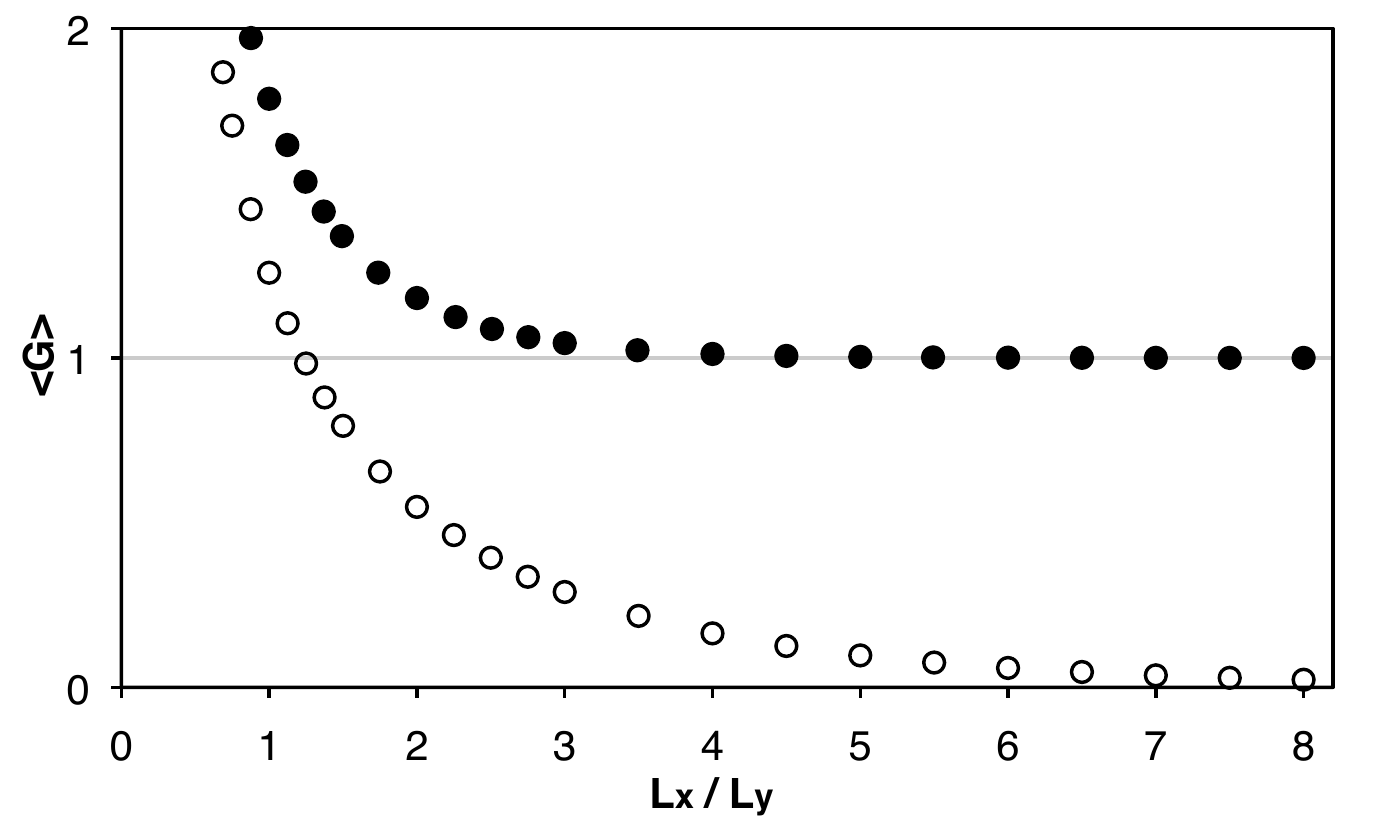}
  \caption{System length dependence of the average conductances at transition points in the symplectic class
           with a PCC at $L_y=65$ ($\bullet$)
           and without a PCC at $L_y=64$ ($\circ$).
           }
  \label{x-dep_S}
 \end{minipage}
\end{figure}

 Figures \ref{x-dep_U} and \ref{x-dep_S} show the dependence on the longitudinal system length $L_x$ of the conductances calculated for the unitary and the symplectic network models, respectively.
 Since the parameters controlling the phases for both models are the same as in the previous section, the conductances at $L_x/L_y=1$ in these figures are consistent with $G\^{2D}$ in Table \ref{G_2D}.
 Because of the metallic property of the PCC, in Q1D limit $L_x/L_y \-> \infty$, the average conductance with a PCC is $\braket<G\^{Q1D}>\-> 1$, 
while that without a PCC is $\braket<G\^{Q1D}>\-> 0$.
 In both classes, the conductance decay length $\lambda$, defined via
	\begin{equation} 
	 \braket<G> = N\_{PCC}+A\exp{\(-\frac{2L_x/L_y}{\lambda}\)},
	\end{equation}
of a network with a PCC is shorter than that of a network without a PCC (Table \ref{G_1D}).


\begin{table}[h] 
 \caption{Critical conductance in a quantum wire.
          Note that the definition of the decay length differs from that of the decay length of the geometric mean of the conductance.
          For example, the decay length estimated from the geometric mean in the symplectic network with $N\_{PCC}=0$ is $1.5$, consistent with ref \cite{ObuseNew}.}
 \label{G_1D}
 \begin{center}
  \begin{tabular}{rcc}
   \br
    universality class & $N\_{PCC}$ & decay length $\lambda$\\
   \mr
    \ unitary     \  &  1  &  1.1 \\
    \ unitary     \  &  0  &  3.2 \\[1mm]
    \ symplectic  \  &  1  &  1.5 \\
    \ symplectic  \  &  0  &  4.0 \\
   \br
  \end{tabular}
 \end{center}
\end{table}

\section{Conclusion}
 In summary, we have studied the transport properties of unitary and symplectic networks with one perfectly conducting channel.
 The system with a PCC has a peculiar critical conductance distribution in 2D.
 Even in the quasi-1D limit, the system remains metallic.
 The PCC suppresses the conduction through the other channels and reduces the conductance decay length.
 A study of the transport properties at the quantum spin Hall transition (metal-topological insulator transition in the symplectic systems without a PCC) will be published elsewhere.

\ack
 This work was supported by Grant-in-Aid No. 18540382.
 H.O. is supported by JSPS Research Fellowships for Young Scientists.

\end{document}